\documentclass[aps,prl,reprint,superscriptaddress]{revtex4-2} 
\usepackage[dvipsnames]{xcolor}
\usepackage[colorlinks=true,urlcolor=blue,citecolor=purple,linkcolor=red]{hyperref}
\usepackage{amssymb,amsmath,amsfonts}
\usepackage{epsfig}
\usepackage{graphicx}
\usepackage{epstopdf}
\usepackage{natbib}
\usepackage{braket}

\def\clock{{\count0=\time
		\divide\count0 60
		\ifnum\count0<10 0\fi\the\count0
		\multiply\count0 -60 \advance\count0 \time
		:\ifnum\count0<10 0\fi \the\count0
}}
\newcommand{\timestamp}{{\small\vbox{\hbox{\tt\jobname.tex}
			\hbox{\the\day/\the\month/\the\year, \clock}}}}

\hypersetup{%
	pdftitle   = {Carrollian fluids and spontaneous breaking of boost symmetry},
	pdfkeywords = {Carrollian hydrodynamics, exotic symmetries, geometry},
	pdfauthor  = {Jay Armas, Emil Have},
}

\newcommand{\C}{\mathbb{C}}

\newcommand{\nn}{\nonumber}

\newcommand{\be}{\begin{eqnarray}}
	\newcommand{\ee}{\end{eqnarray}}
\newcommand{\beq}{\begin{eqnarray}}
	\newcommand{\eeq}{\end{eqnarray}}

\newcommand{\beqa}{\begin{eqnarray}}
	\newcommand{\eeqa}{\end{eqnarray}}
\newcommand{\D}{{\partial}}

\newcommand{\pd}[2]{\frac{\partial #1}{\partial #2}}

\usepackage{mathrsfs}

\newcommand{\h}{\mathfrak{h}}

\newcommand{\m}{\mathfrak{m}}

\newcommand{\verts}[1]{\left\vert #1 \right\vert}

\ifdefined\C
\renewcommand{\C}{\boldsymbol{C}}
\else
\newcommand{\C}{\boldsymbol{C}}
\fi

\definecolor{gris}{rgb}{0.5,0.5,0.5}
\definecolor{darkgreen}{rgb}{0.0,0.5,0.0}

\allowdisplaybreaks[0]
\begin{document}

	\title{Carrollian fluids and spontaneous breaking of boost symmetry}

	\author{Jay Armas}
	\email{j.armas@uva.nl}
	\affiliation{Institute for Theoretical Physics and Dutch Institute for Emergent Phenomena, University of Amsterdam, 1090 GL Amsterdam, The Netherlands}
	\affiliation{Institute for Advanced Study, University of Amsterdam, Oude Turfmarkt 147, 1012 GC Amsterdam, The Netherlands}

	\author{Emil Have}
	\email{emil.have@nbi.ku.dk}
	\affiliation{School of Mathematics and Maxwell Institute for Mathematical Sciences,\\
		University of Edinburgh, Peter Guthrie Tait road, Edinburgh EH9 3FD, UK } 
	\affiliation{Niels Bohr International Academy, Niels Bohr Institute, University of Copenhagen, Blegdamsvej 17, DK-2100 Copenhagen Ø, Denmark}

	\begin{abstract}
		In the hydrodynamic regime, field theories typically have their boost symmetry spontaneously broken due to the presence of a thermal rest frame although the associated Goldstone field does not acquire independent dynamics. We show that this is not the case for Carrollian field theories where the boost Goldstone field plays a central role. This allows us to give a first-principles derivation of the equilibrium currents and dissipative effects of Carrollian fluids. We also demonstrate that the limit of vanishing speed of light of relativistic fluids is a special case of this class of Carrollian fluids. Our results shine light on the thermodynamic properties and thermal partition functions of Carrollian field theories.
	\end{abstract}

	\maketitle
	\textbf{Introduction.} In the past few years Carrollian physics, emerging by taking the limit of vanishing speed of light, has been found useful for describing a variety of phenomena in contexts ranging from black holes~\cite{Penna:2018gfx,Donnay:2019jiz, Hansen:2021fxi,Redondo-Yuste:2022czg}, cosmology~\cite{deBoer:2021jej, Bagchi:2021qfe}, gravity \cite{Hartong:2015xda,Bergshoeff:2017btm,Ciambelli:2018ojf,Perez:2021abf,Guerrieri:2021cdz,Hansen:2021fxi,Henneaux:2021yzg,Figueroa-OFarrill:2022mcy,Campoleoni:2022ebj,Fuentealba:2022gdx,Perez:2022jpr}, to hydrodynamics~\cite{Ciambelli:2018wre,Redondo-Yuste:2022czg, Ciambelli:2018xat, Campoleoni:2018ltl, deBoer:2021jej, Petkou:2022bmz, Freidel:2022bai, Freidel:2022vjq, Bagchi:2023ysc, deBoer:2023fnj}. Concretely, Carrollian fluids can be used to describe Bjorken flow, which is relevant for models of the quark-gluon plasma, cf.~\cite{Bagchi:2023ysc} (and its conformal generalisation, Gubser flow~\cite{Bagchi:2023rwd}). Carrollian fluids also model dark energy in inflationary models~\cite{deBoer:2021jej}. Furthermore, Carrollian symmetries are expected to have a role to play in exotic phases of matter (e.g., via Carroll/fracton dualities \cite{Bidussi:2021nmp, Marsot:2022imf, Figueroa-OFarrill:2023vbj,Figueroa-OFarrill:2023qty} and in superconducting twisted bilayer graphene \cite{Bagchi:2022eui}). 
	
	Many of the properties encountered in this Carrollian limit are expected to be explained from underlying quantum field theories with inherent Carrollian symmetries. Indeed, if conformal symmetry is present in addition, such theories would be putative holographic duals to flat space gravity \cite{Duval:2014uva,Duval:2014lpa,Bagchi:2016bcd,Figueroa-OFarrill:2021sxz,Donnay:2022aba,Donnay:2022wvx,Bagchi:2022emh,Baiguera:2022lsw,Bagchi:2023fbj,Saha:2023hsl,Saha:2023abr}. However, when attempting to formulate such Carrollian field theories, several issues have been pointed out including violations of causality, lack of well-defined thermodynamics and ill-defined partition functions \cite{deBoer:2021jej, deBoer:2023fnj}. Our goal is to show that the lack of well-defined thermodynamics in Carrollian field theories is expected in the hydrodynamic regime but that this issue can be cured when carefully accounting for the Carrollian symmetries. 
	
	The approach we take is to consider the hydrodynamic regime of such putative Carrollian field theories and show how to construct their equilibrium partition function and near-equilibrium dynamics. In particular, we will show that there is no proper notion of temperature in Carrollian fluids unless the Goldstone field of spontaneous broken boost symmetry is taken into account. This allows us to construct a well-defined hydrodynamic theory of Carrollian fluids (similar to \textit{framids} in the language of \cite{Nicolis:2015sra, Alberte:2020eil}). In the process, we show that seemingly different approaches to Carrollian hydrodynamics previously pursued in the literature \cite{Ciambelli:2018wre, Ciambelli:2018xat, Campoleoni:2018ltl, deBoer:2021jej, Petkou:2022bmz, Freidel:2022bai, Freidel:2022vjq, Bagchi:2023ysc, deBoer:2023fnj} are in fact equivalent and special cases of the Carrollian fluids we derive.
	
	The fact that boost symmetry is spontaneously broken in hydrodynamics is not unexpected. Thermal states break the boost symmetry spontaneously due to the presence of a preferred rest frame aligned with the thermal vector \cite{Alberte:2020eil, Komargodski:2021zzy}. In the context of hydrodynamics the thermal vector is the combination $u^\mu/T$ of the unit normalised fluid velocity $u^\mu$ and temperature $T$. On the other hand, the Goldstone field associated with the breaking of boost symmetry does not typically feature in the low energy spectrum of the theory because it is determined in terms of the other dynamical fields (see, e.g., \cite{Nicolis:2015sra}). This is easy to show for relativistic fluids. Consider a $(d+1)$-dimensional spacetime metric $g_{\mu\nu}={E_\mu^A} {E^\mu_A}$ where ${E_\mu^A}$ is the set of vielbeins, $\mu=0,\dots,d$ are spacetime indices and $A=0,\dots,d$ are internal Lorentz indices. The Goldstone field associated to spontaneous breaking of Lorentz boost symmetry is the vector $\ell_A={\Lambda_A}^0$ where $\Lambda_A{}^0$ is the Lorentz boost matrix \cite{Nicolis:2015sra}, and acquires an expectation value $\langle \ell_A\rangle=\delta_A^0$ in the ground state. In thermal equilibrium we can construct an equilibrium partition function $S=\int d^{d+1}x\sqrt{-g} P(T,u^\mu \ell_\mu)$ where $P$ is the fluid pressure. The temperature is given by $T=T_0/|K|$ with $T_0$ a global constant temperature, and the fluid velocity is $u^\mu=K^\mu/|K|$ satisfying $u_\mu u^\mu=-1$ and defined in terms of the Killing (thermal) vector $K^\mu$ with modulus $|K|^2=-g_{\mu\nu}K^\mu K^\nu$. We furthermore defined $\ell_\mu={E_\mu^A}\ell_A$ satisfying $\ell_\mu\ell^\mu=-1$. The Goldstone equation of motion obtained from varying the partition function $S$ with respect to $\ell_A$ is \cite{Nicolis:2015sra}
	\begin{equation}\label{eq:relGoldstone}
		\left(\delta^\mu_\nu +\ell^\mu \ell_\nu\right)\frac{\delta S}{\delta \ell_\mu}=0\,.
	\end{equation}
	This equation implies $u^\mu=-\ell_\nu u^\nu \ell^\mu$ for arbitrary thermodynamic coefficients, and can only be satisfied if $\ell^\mu=u^\mu$. Indeed, we see that the dynamics of the Goldstone field $\ell^\mu$ is determined by the dynamics of the fluid velocity and hence can be removed from the hydrodynamic description. The same conclusion is reached for the case of spontaneous breaking of Galilean (Bargmann) boost symmetry \cite{Glorioso:2023chm} (see also \cite{Komargodski:2021zzy}). However, in the case of Carrollian symmetry, as we will show, the boost Goldstone acquires its own independent dynamics. Below we introduce Carrollian geometry and use it to show that, naively, there is no well-defined notion of temperature.\\

	\textbf{Carrollian geometry and the lack of temperature.} A \textit{weak} Carrollian geometry on a $(d+1)$-dimensional manifold $M$ is defined by a Carrollian structure $(v^\mu,h_{\mu\nu})$ consisting of the nowhere-vanishing Carrollian vector field $v^\mu$, and the corank-1 symmetric tensor $h_{\mu\nu}$, the ``ruler'', satisfying $h_{\mu\nu} v^\nu=0$. It is useful to define inverses $(\tau_\mu, h^{\mu\nu})$ satisfying $v^\mu \tau_\mu = -1$, and $\tau_\mu h^{\mu\nu} =0$, as well as the completeness relation $-v^\mu \tau_\nu + h^{\mu\rho}h_{\rho\nu} = \delta_\nu^\mu$. It is also useful to introduce the spatial vielbeins $e^a_{\mu}$ and their inverses $e_a^{\mu}$ which can be used to write $h_{\mu\nu}=e^a_{\mu}e_{a\nu}$. Under Carrollian boosts, the inverses transform as
	\begin{equation}
		\delta_C \tau_\mu = \lambda_\mu\,,\qquad \delta_C h^{\mu\nu} = 2\lambda_\rho h^{\rho(\mu}v^{\nu)}\,,
	\end{equation}
	corresponding to $\delta_C e^\mu_a =v^\mu\lambda_a$, where we have used that $v^\mu\lambda_\mu=0$. A \textit{strong} Carrollian geometry is a {weak} Carrollian geometry together with an affine connection. We focus mainly on {weak} Carrollian geometry but discuss {strong} Carrollian geometry in Appendix~\hyperref[app:strong-fluids]{D}.
	
	Given the Carroll geometry and the existence of a thermal vector in equilibrium, namely the spacetime Killing vector $k^\mu$, one can proceed as for other (non)-Lorentzian field theories~\cite{PartitionFunction2, Banerjee:2012iz, Armas:2013hsa, Jensen:2014ama, Banerjee:2015uta, deBoer:2017ing,deBoer:2017abi, Armas:2019gnb, Jain:2020vgc, Armas:2020mpr, Armas:2023ouk, Jain:2023nbf} and construct an equilibrium partition function by identifying the invariant scalars under all local symmetries which the pressure $P$ can depend on, as above Eq.~\eqref{eq:relGoldstone} for the relativistic case. A more thorough construction of the partition function will be given in a later section. Here we note that for non-relativistic theories the temperature $T$ is given by the scalar $T=T_0/(k^\mu\tau_\mu)$ with $T_0$ a constant global temperature. However, in the Carrollian case $T$ is not invariant under boost transformations since $\delta_C(k^\mu\tau_\mu)=k^\mu\lambda_\mu\ne0$. Indeed, there is no well-defined notion of temperature for arbitrary observers \footnote{It is possible to construct an invariant scalar, namely $h_{\mu\nu}k^\mu k^\nu$. However, this leads to a theory in which different observers have different notions of temperature. This theory appeared in~\cite{deBoer:2023fnj}, although it remains unclear to what extent it could be interpreted in terms of a theory of fluids.}. This is rooted in the fact that it is not possible to impose a timelike normalisation condition on spacetime vectors such as the fluid velocity since $u^\mu\tau_\mu$ is not boost invariant. This  argument does not rely on any specific model of Carrollian (quantum) field theory since these statements are valid in the hydrodynamic regime. This suggests that well-defined thermodynamic limits of such putative theories are subtle. Below we introduce the boost Goldstone and use it to show that it can be used to define an appropriate notion of temperature.\\

	\textbf{The Carroll boost Goldstone.} We define the boost Goldstone as the vector $\theta^\mu$ which transforms under Carrollian boosts $\lambda_\mu$ as
	\begin{equation}
		\label{eq:trafo-vector-Goldstone}
		\delta_C \theta^\mu = -h^{\mu\nu}\lambda_\nu\,,
	\end{equation}
	where $\lambda_\mu v^\mu = 0$. This implies that only the spatial part of $\theta^\mu$ is physical, which we can enforce by endowing the Goldstone with a timelike Stueckelberg symmetry of the form
	\begin{equation}
		\label{eq:Stueckelberg-sym}
		\delta_S \theta^\mu = \chi v^\mu\,,
	\end{equation}
	where $\chi$ is an arbitrary function \footnote{Alternatively, we could have introduced a Goldstone $\theta_a$, where $a$ is a tangent space index, but that would require us to work in a first order formalism.}. With this we can build the boost and Stueckelberg invariant vielbeine
	\begin{equation}
		\label{eq:inv-fields}
		\begin{split}
			\hat \tau_\mu = \tau_\mu + h_{\mu\nu} \theta^\nu\,,\qquad 
			\hat e^\mu_a = e^\mu_a + v^\mu \theta^\mu e_{\mu a} \,,
		\end{split}
	\end{equation}
	which lead to the following invariant ruler
	\begin{equation}
		\hat h^{\mu\nu} = \delta^{ab}\hat e^\mu_a \hat e^\nu_b = h^{\mu\nu} + v^\mu v^\nu (\theta^2 + 2\tau_\rho \theta^\rho) + 2 v^{(\mu}\theta^{\nu)}\,,
	\end{equation}
	where $\theta^2 = h_{\mu\nu}\theta^\mu\theta^\nu$. 
	Together with $v^\mu$ and $h_{\mu\nu}$, these form an \textit{Aristotelian} structure~\footnote{Note that we also have the boost invariant combination $\bar h_{\mu\nu} := h^{\mu\nu} + 2v^{(\mu}\theta^{\nu)}$, but it does not form part of an Aristotelian structure and it is not invariant under Stueckelberg transformations. The effective geometry obtained differs from that in \cite{Hartong:2015xda} as the time component of the boost Goldstone is not physical and $\theta^\mu$ is not a background field.}
	\begin{equation}
		\hat h^{\mu\nu}\hat \tau_\nu = 0~,~ v^\mu \hat \tau_\mu = -1~,~ \hat h^{\mu\rho}h_{\rho\nu} = \delta^\mu_\nu + v^\mu \hat\tau_\nu =: \hat h^\mu_\nu\,,
	\end{equation}
	that is partly dynamical due to the Goldstone $\theta^\mu$. The low-energy effective action for the Carrollian boost Goldstone is a two-derivative Ho\v rava--Lifshitz type action as we show in Appendix \hyperref[app:HL]{A}. If coupled to Carrollian gravity the resultant action would be derivable from the limit of vanishing speed of light of Einstein--Aether theory \cite{Jacobson:2000xp}. Before showing how the Goldstone allows to define a notion of temperature, we first discuss the currents and conservation laws.\\
	
	\textbf{Currents and conservation laws.} We now consider an arbitrary fluid functional (or free energy) $S[\tau_\mu,h_{\mu\nu};\theta^\nu]$ for a Carrollian fluid with spontaneously broken boosts. The variation of this functional is
	\begin{equation}
		\label{eq:fluid-functional-variation}
		\begin{split}
			\delta S &= \int d^{d+1}x\,e\left[ - T^\mu \delta \tau_\mu + \frac{1}{2}\mathcal{T}^{\mu\nu}\delta h_{\mu\nu} - K_\mu \delta \theta^\mu\right]\,,
		\end{split}
	\end{equation}
	where $T^\mu$ is the energy current, $\mathcal{T}^{\mu\nu}$ the stress-momentum tensor and $K_\mu$ the response to the Goldstone field. In particular $K_\mu=0$ gives the analogous equation of motion for the Goldstone as in \eqref{eq:relGoldstone}.
	The measure is defined as $e = \det(\tau_\mu,e_\mu^a) = \det(\hat\tau_\mu,e_\mu^a)$. The Ward identities for the Stueckelberg and boost symmetries are, respectively,
	\begin{equation}
		\label{eq:WIs}
		v^\mu K_\mu = 0\,,\qquad T^\nu h_{\nu\mu} = K_\mu\,.
	\end{equation}
	The equation of motion for the Goldstone, $K_\mu=0$, imposes the condition $T^\nu h_{\nu\mu} = 0$. In other words, the boost Ward identity now becomes the equation of motion for the Goldstone. The momentum-stress tensor is not boost invariant. In fact, computing the second variation, which must vanish $\delta_{C}( \delta_C S ) = 0$, we find that $\delta_C T^\mu =\delta_C K_\mu = 0$ and $\delta_C \mathcal{T}^{\mu\nu} = 2T^{(\mu}h^{\nu)\rho}\lambda_\rho$. The associated energy-momentum tensor (EMT) $T^\mu{_\nu}= -\tau_\nu T^\mu + \mathcal{T}^{\mu\rho}h_{\rho\nu}$ is also not boost invariant and transforms as
	\begin{equation}
		\label{eq:EMT-boost-trafo}
		\delta_C T^\mu{_\nu} =  K_\nu h^{\mu\rho} \lambda_\rho \,,
	\end{equation}
	where we used~\eqref{eq:WIs}. Doing the same for the Stueckelberg symmetry, the condition $\delta_{S}( \delta_S S ) = 0$ implies that $\delta_S T^\mu{_\nu} = -\chi v^\mu K_\nu$. Hence, the EMT is both boost and Stueckelberg invariant if $K_\mu = 0$. The diffeomorphism Ward identity reads
	\begin{equation}
		\label{eq:diffeo-WI-Carroll}
		\begin{split}
			e^{-1}\D_\mu(e  T^\mu{_\rho}) + T^\mu \D_\rho \tau_\mu - \frac{1}{2}{\mathcal{T}}^{\mu\nu}\D_\rho h_{\mu\nu}=0\,,
		\end{split}
	\end{equation}
	where we used that $K_\mu=0$. It is possible to obtain manifestly boost invariant currents, including the EMT, by formulating the action in terms of the effective Aristotelian structure~\eqref{eq:inv-fields}, as we show in Appendix \hyperref[app:dissipation]{B}.\\
	
	\textbf{Equilibrium partition function and Carrollian fluids.} To derive the currents of Carrollian fluids, we consider the equilibrium partition function construction. An equilibrium Carrollian background consists of a set of symmetry parameters $K = (k^\mu, \lambda^\mu_K,\chi_K)$, where $k^\mu$ is a Killing vector and $\lambda_K^\mu$ is a boost symmetry parameter, while $\chi_K$ is a Stueckelberg symmetry parameter. The various structures transform according to
	\begin{align}
		\delta_K v^\mu &= \pounds_k v^\mu = 0\,,\qquad\delta_K\tau_\mu = \pounds_k \tau_\mu + \lambda_\mu^K =0\,,\nn\\
		\delta_K h_{\mu\nu} &= \pounds_k h_{\mu\nu}=0\,,\nn\\  \delta_K h^{\mu\nu} &=\pounds_k h^{\mu\nu} + 2\lambda_\rho^{K} h^{\rho(\mu}v^{\nu)}=0\,,\nn\\
		\delta_K \theta^\mu &= \pounds_k \theta^\mu - h^{\mu\nu}\lambda^K_\mu + \chi^K v^\mu = 0 \,.\label{eq:equil-transformations}
	\end{align}
	The boost and Stueckelberg symmetry parameters transform as 
	\begin{equation} \label{eq:invariants}
		\delta \lambda_\mu^K = \pounds_\xi \lambda_\mu^K - \pounds_k \lambda_\mu\,,\qquad \delta \chi^K = \pounds_\xi \chi^K - \pounds_k \chi\,,
	\end{equation}
	under infinitesimal diffeomorphisms generated by $\xi^\mu$, infinitesimal Carrollian boosts $\lambda_\mu$ and Stueckelberg transformations $\chi$. As we show in Appendix \hyperref[app:dissipation]{B}, $\lambda_\mu^K$ and $\chi^K$ will not play a role in the effective fluid description. 
	
	Before enumerating the possible invariant scalars, we must provide a gradient ordering. As usual we take the geometry itself to be of ideal order, that is, $\tau_\mu\sim h_{\mu\nu}\sim v^\mu\sim h^{\mu\nu}\sim \mathcal{O}(1)$. Since $\theta^\mu$ enters the definition of $\hat \tau_\mu$ it must have the same ordering, $\theta^\mu\sim\mathcal{O}(1)$. Gradients of these structures are $\mathcal{O}(\partial)$ and hence suppressed in a hydrodynamic expansion. Given this gradient scheme the only two ideal order invariants are
	\begin{equation}
		T = T_0/\hat\tau_\mu k^\mu\,,\qquad \vec u^2 = h_{\mu\nu} u^\mu u^\nu\,,
	\end{equation}
	where $u^\mu = k^\mu/\hat\tau_\rho k^\rho$, which satisfies $\hat\tau_\mu u^\mu = 1$. We note that we can now define a notion of temperature $T$ that is invariant for all observers. The scalar $\vec u^2$ is the modulus of the spatial fluid velocity. Generically the fluid velocity can be decomposed as  $u^\mu = -v^\mu + \vec u^\mu$, where $\vec u^\mu = \hat h^\mu_\nu u^\nu$ with $\hat h^\mu_\nu := \hat h^{\mu\rho}h_{\rho\nu}$. We furthermore define $\vec u_\mu = h_{\mu\nu} u^\nu = h_{\mu\nu}\vec u^\nu$, such that $\vec u^\mu = \hat h^\mu_\nu u^\nu$, but $\vec u^\mu \neq h^{\mu\nu} \vec u_\nu$. Note in particular that $u^\mu$ decomposes as follows relative to the Carrollian structure
	\begin{equation}
		\label{eq:u-relative-to-Carroll}
		u^\mu = -v^\mu(1 - \theta^\nu \vec u_\nu) + h^\mu_\nu u^\nu\,.
	\end{equation}
	The hydrostatic partition function at ideal order is given by $S = \int d^{d+1}x\,eP(T,\vec u^2)$. Using the general action variation \eqref{eq:fluid-functional-variation} together with the ``variational calculi'' $\delta h^{\mu\nu} = 2v^{(\mu} h^{\nu)\rho}\delta\tau_\rho - h^{\mu\rho}h^{\nu\sigma}\delta h_{\rho\sigma}$ and $\delta v^\mu = v^\mu v^\nu \delta \tau_\nu - h^{\mu\nu} v^\rho \delta h_{\rho\nu}$ we obtain the ideal order currents
	\begin{equation} \label{eq:CarrollStress}
		\begin{split}
			T_{(0)}^\mu &= Pv^\mu + s T u^\mu + m \vec u^2 u^\mu\,,\\
			\mathcal{T}_{(0)}^{\mu\nu} &= P h^{\mu\nu} + m u^\mu u^\nu - 2(sT + m\vec u^2) u^{(\mu}\theta^{\nu)}\,,\\
			K_{(0)\mu} &= (sT + m\vec u^2) \vec u_\mu\,,
		\end{split}
	\end{equation}
	where the subscript $(0)$ indicates that the currents are of ideal order $\mathcal{O}(1)$ and the entropy $s$ and mass density $m$ are defined via $dP=sdT+md\vec u^2$. The associated EMT is given by
	\begin{equation}
		\label{eq:ideal-EMT}
		\hspace{-0.1cm}
		\begin{split}
			T^\mu_{(0)\nu} &= P \delta^\mu_\nu + m u^\mu \vec u_\nu - (sT + m \vec u^2)(u^\mu\hat \tau_\nu + \theta^\mu \vec u_\nu)\,,
		\end{split}
	\end{equation}
	which transforms as in \eqref{eq:EMT-boost-trafo}. 
	
	The equation of motion for the Goldstone $K_{(0)\mu}=0$, which is equivalent to the boost Ward identity, gives a constraint on the dynamics
	\begin{equation}
		\label{eq:boost-WI-ideal}
		\left( sT + m \vec u^2 \right) \vec u_\mu = 0\,,
	\end{equation}
	and can be viewed as a framid condition for Carrollian fluids. Defining the energy density as $\mathcal{E} = \hat\tau_\mu T^\mu_{(0)}$, the Goldstone equation has two branches of solutions: either $\mathcal{E} + P = sT+m\vec u^2= 0$ or $\vec u_\mu=0$. Neither of them allows for the elimination of the Goldstone $\theta^\mu$ from the low-energy description. The constraint~\eqref{eq:boost-WI-ideal} was derived in equilibrium, but we show in Appendix~\hyperref[app:dissipation]{B} that it also holds off-equilibrium, although it receives corrections due to dissipative effects. As such, together with \eqref{eq:diffeo-WI-Carroll}, it provides the ideal order dynamics for Carrollian fluids. Eqs.~\eqref{eq:CarrollStress}--\eqref{eq:boost-WI-ideal} are a central result of this work as they provide a well-defined notion of Carrollian fluids. Below we show that the $c\rightarrow 0$ limit of relativistic fluids gives rise to a Carrollian fluid with $\vec u_\mu=0$. \\

	\textbf{The $c\rightarrow 0$ limit of a relativistic fluid.} The $c\rightarrow 0$ limit of relativistic fluids was considered in~\cite{Ciambelli:2018wre,Ciambelli:2018xat,Petkou:2022bmz} (see also~\cite{Bagchi:2023ysc}) for a specific class of metrics. The same limit was taken in~\cite{deBoer:2023fnj}, where it was referred to as a ``timelike fluid''. Here, we demonstrate that these notions coincide and correspond to the special case of the Carrollian fluid we introduced above with $\vec u_\mu = 0$, and that the emergence of the Goldstone can be understood from the ultra-local expansion of the Lorentzian geometry. The relativistic EMT is given by
	\begin{equation}
		\label{eq:relativistic-EMT}
		{T^\mu}_\nu=\frac{\hat{{\mathcal{E}}}+\hat P}{c^2}U^\mu U_\nu+\hat P\delta^{\mu}_\nu\,, 
	\end{equation}
	where $U^\mu U^\nu g_{\mu\nu} = -c^2$, and where the ``hat'' indicates relativistic thermodynamic quantities. To take the limit, we first consider the metric and its inverse in ``pre-ultralocal (PUL) variables''~\cite{Hansen:2021fxi}
	\begin{equation}
		g_{\mu\nu} = -c^2 T_\mu T_\nu + \Pi_{\mu\nu}\,,\quad g^{\mu\nu} = - \frac{1}{c^2}V^\mu V^\nu + \Pi^{\mu\nu}\,,
	\end{equation}
	where $T_\mu V^\mu = -1$, $T_\mu \Pi^{\mu\nu} = V^\mu \Pi_{\mu\nu} = 0$, $\Pi^{\mu\rho}\Pi_{\rho\nu} = \delta^\mu_\nu + V^\mu T_\nu$. The leading order components of the PUL variables correspond to the fields that make up the Carrollian structure, e.g., $V^\mu = v^\mu + \mathcal{O}(c^2)$. We write the expansion of the relativistic fluid velocity relative to the PUL variables as
	\begin{equation*}
		U^\mu = -V^\mu - c^2 \mathfrak u^\mu \,, 
	\end{equation*}
	for some $\mathfrak u^\mu$. Crucially, $U^\mu$ is invariant under local Lorentz boosts while $\delta_C V^\mu = c^2 h^{\mu\nu}\lambda_\nu + \mathcal{O}(c^4)$, implying that $\delta_C \mathfrak u^\mu = -h^{\mu\nu}\lambda_\nu + \mathcal{O}(c^2)$. This shows that $\mathfrak u^\mu$ cannot be identified with a fluid velocity in the Carrollian limit. Indeed,  we may identify the spatial part of the leading order term in the $c^2$-expansion of $\mathfrak u^\mu$ with the spatial part of the boost Goldstone
	\begin{equation}
		\Pi_{\mu\nu}\mathfrak u^\nu = h_{\mu\nu}\theta^\nu + \mathcal{O}(c^2) =: \vec\theta_\mu + \mathcal{O}(c^2)\,.
	\end{equation}
	Using this, together with $U_\mu=c^2\hat\tau_\mu+\mathcal{O}(c^4)$ the EMT becomes
	\begin{equation}
		\label{eq:EMT-from-limit}
		T^\mu{_\nu} = (\mathcal{E} + P) v^\mu \hat \tau_\nu + P\delta^\mu_\nu + \mathcal{O}(c^2)\,,
	\end{equation}
	where $\mathcal{E}$ and $P$ are the leading order contributions of $\hat{\mathcal{E}}$ and $\hat P$, respectively, satisfying the Euler relation $\mathcal{E} + P = sT$. This is exactly the ``timelike'' fluid of~\cite{deBoer:2023fnj}, corresponding to the $\vec u_\mu=0$ branch of the Carrollian fluid we described above~\footnote{The Lagrange multiplier $\chi^\mu$ used in the hydrostatic partition function construction of the ``timelike'' fluid in~\cite{deBoer:2023fnj} can be identified with the Goldstone field $\theta^\mu$ up to a suitable redefinition.}.
	
	It is instructive to take the $c\to0$ limit of the relativistic equation of motion
	$\hat\nabla_\mu T^\mu{_\nu} = 0$, where $\hat\nabla$ is the Levi-Civita connection of the spacetime metric $g_{\mu\nu}$. Deferring the details to Appendix \hyperref[app:limit]{C}, we note here that the equations of motion in the limit $c\rightarrow 0$ can be expressed as
	\begin{equation}
		\label{eq:limit-equations}
		\begin{split}
			v^\mu \D_\mu {\mathcal{E}} &= ({\mathcal{E}} + P) K\,,\\
			h^{\mu\nu}\D_\nu P &= -\widetilde\varphi^\mu ({\mathcal{E}} + P) + ({\mathcal{E}} + P)K h^{\mu\nu}\vec\theta_\nu \\
			&\quad- h^{\mu\nu} v^\rho \tilde\nabla_\rho[({\mathcal{E}} + P)\vec\theta_\nu]\,,
		\end{split}
	\end{equation}
	where $K = h^{\mu\nu}K_{\mu\nu}:=-\frac{1}{2}h^{\mu\nu}\pounds_v h_{\mu\nu}$ is the trace of the intrinsic torsion of the Carrollian structure~\cite{Figueroa-OFarrill:2020gpr}, and is sometimes referred to as the ``Carrollian expansion'', while  $\widetilde\varphi^\lambda = 2h^{\lambda \mu}v^\nu \D_{[\nu}\tau_{\mu]} - h^{\lambda\mu}h^{\nu\sigma}\vec\theta_\sigma K_{\mu\nu}$. The first term in $\widetilde \varphi^\lambda$ is sometimes referred to as the ``Carrollian acceleration'', while the second term comes from the $c^2$ expansion of the Levi-Civita connection. The covariant derivative $\tilde\nabla$ is a Carroll compatible connection that arises in the $\mathcal{O}(1)$ piece of the $c^2$ expansion of the Levi-Civita connection, which we discuss further in Appendix~\hyperref[app:limit]{C}. These equations are fully covariant and reduce to the special case of the equations of motion obtained in~\cite{Ciambelli:2018wre,Ciambelli:2018xat,Petkou:2022bmz,Bagchi:2023ysc} when restricted to spacetime metrics that admit a Randers--Papapetrou parameterisation. Furthermore, these equations can be obtained by projecting the conservation law \eqref{eq:diffeo-WI-Carroll} along the time and spatial directions using \eqref{eq:ideal-EMT} with $\vec u_\mu=0$. We thus have shown that the ``timelike'' fluid of \cite{deBoer:2023fnj} is the same as the Carrollian fluid of \cite{Ciambelli:2018wre,Ciambelli:2018xat,Bagchi:2023ysc}, and both are a special case of the Carrollian fluid derived here.\\

	\textbf{Dissipation and modes.} In Appendix~\hyperref[app:dissipation]{B} we show that at order $\mathcal{O}(\partial)$ the class of Carrollian fluids we introduced is characterised by 2 hydrostatic coefficients and 10 dissipative coefficients. Here we study the effect of specific coefficients in the linear spectrum of fluctuations. We consider flat Carrollian space with  $\tau_\mu = \delta_\mu^t$, $ v^\mu = -\delta^\mu_t$, $h_{\mu\nu} = \delta_\mu^i \delta_\nu^i$ and 
	$ h^{\mu\nu} = \delta^\mu_i \delta^\nu_i$ (see Appendix~\hyperref[app:dissipation]{B} for more details). We then fluctuate the conservation equations \eqref{eq:diffeo-WI-Carroll} and the boost Ward identity \eqref{eq:WIs} around an equilibrium state with constant temperature $T_0$, fluid velocity $v_0^i$ and Goldstone field $\theta_0^i$, such that, e.g., $\theta^i=\theta_0^i+\delta\theta^i$. Using plane wave perturbations with frequency $\omega$ and wave-vector $\vec k$ we find a distinguishing feature of these Carrollian fluids: the mode structure strongly depends on whether the equilibrium state carries nonzero velocity $v_0^i$. If $v_0^i = 0$, the linearised equations only admit a nontrivial solution if $\theta_0^i\neq 0$, $T_0 = 0$ and $\delta v^i = 0$. Denoting the angle between the momentum $k_i$ and $\theta_0^i$ by $\phi$, this leads to a single linear mode 
	\begin{equation}
		\omega = -\frac{1}{\vert\vec\theta_0\vert\cos\phi} \vert\vec k\vert\,,
	\end{equation}
	where we assumed that the value of the entropy density $s$ in equilibrium $s_0$ remains finite and non-vanishing when $T_0\to0$, otherwise there is no mode. Interestingly, this mode is not affected by any of the 12 transport coefficients entering at order $\mathcal{O}(\partial)$~\footnote{This statement relies on the assumption that all transport transport coefficients remain finite and non-vanishing for $T_0 \to 0$ and $v_0^i \to 0$.}. This spectrum corresponds to the branch of solutions with $\vec u_\mu=0$ and hence it is the expected spectrum arising from the $c\to0$ limit of an ideal relativistic fluid. 
	
	On the other hand if $v_0^i\neq 0$ but $\theta^i_0 = 0$ a more interesting spectrum can be obtained. For simplicity we only consider the effect of a bulk viscosity $s_3$ and one anisotropic viscosity $s_2$. Besides a gapped mode, we     find for $d=2$ a single diffusive mode of the form 
	\begin{equation} \label{eq:mode2}
		\omega - v^i_0 k_i =
		-\frac{i\Gamma_1}{2} (\varepsilon_{ij} v_0^i  k^j)^2\,,
	\end{equation}
	where $\varepsilon_{ij}$ is the two-dimensional Levi-Civita symbol, and 
	\begin{align*}
		\Gamma_1 &= s_{2,0}\frac{(T_0^2 \chi_{TT} + \verts{v_0}^2(3T_0\chi_{Tu} + 2\verts{v_0}^2\chi_{uu}))}{s_0 T_0 (T_0^2 \chi_{TT} + \verts{v_0}^2(2T_0\chi_{Tu} + \verts{v_0}^2\chi_{uu})) }\\
		&\quad + \frac{2s_{3,0}}{s_0 T_0}\,,
	\end{align*}
	where $s_{3,0}$ is the value of $s_3$ in equilibrium, ditto $s_2$, and where we defined $\chi_{TT} = \left( \pd{^2P}{T\D T} \right)_0$, $\chi_{uu} = \left( \pd{^2P}{\vec{u}^2\D \vec{u}^2} \right)_0$ and $\chi_{Tu} = \left( \pd{^2P}{T\D \vec{u}^2} \right)_0$. The left hand side of \eqref{eq:mode2} is characteristic of a fluid without boost symmetry \cite{Armas:2020mpr} while the right hand side is typical of a diffusive mode. A salient signature of Carrollian fluids is that the spectrum is only non-trivial for states with non-zero equilibrium velocity $v_0^i$.\\

	\textbf{Discussion}
	We have given a first-principles derivation of Carrollian fluids based on symmetries, showing that the spontaneous breaking of boost symmetry plays a crucial role in defining equilibrium partition functions of Carrollian field theories in the hydrodynamic regime. It is interesting to speculate whether this peculiar feature of Carrollian hydrodynamics can shed light on how to construct well-defined partition functions using specific microscopic models of Carrollian field theories along the lines of \cite{Komargodski:2021zzy, deBoer:2023fnj}. 
	
	Different approaches to Carrollian hydrodynamics have appeared in the literature in the past few years \cite{Ciambelli:2018wre, Ciambelli:2018xat, Campoleoni:2018ltl, deBoer:2021jej, Freidel:2022bai, Freidel:2022vjq, Bagchi:2023ysc, deBoer:2023fnj}. Revisiting the $c\to0$ limit of relativistic fluids we showed that there are subtleties regarding the interpretation of the dynamical variables that appear in the limit of the equations of motion. In particular, we showed that what naively appeared to be a fluid velocity was in fact a Goldstone field associated to the spontaneous breaking of boost symmetry. This allowed us to show that the different approaches are not only equivalent to each other but also special cases of the more general Carrollian fluids we introduced here. We believe it could be interesting to revisit the black hole membrane paradigm \cite{Donnay:2019jiz} in light of this new understanding.  
	
	The effective field theory geometry becomes Aristotelian once taking the Goldstone field into account. This allowed us to easily understand the dissipative structure of such fluids using earlier results \cite{deBoer:2020xlc, Armas:2020mpr}. The spectrum of excitations for certain classes of equilibrium states shares certain similarities with the spectrum of excitations of $p$-wave fracton superfluids in which the Goldstone field associated to the spontaneous breaking of dipole symmetry plays an analogous role to the boost Goldstone field \cite{Glorioso:2023chm, Armas:2023ouk, Jain:2023nbf}, albeit in the Carroll case a non-vanishing fluid velocity is needed. We believe that this relation can be made clearer if we consider \textit{strong} Carrollian geometries as we describe in Appendix~\hyperref[app:strong-fluids]{D}.
	
	Finally, it would be interesting to consider the addition of conformal symmetry as this could shine light on thermodynamic properties of holographic dual theories of flat space gravity. If we impose this symmetry the equation of state for the branch $\vec u_\mu=0$ becomes $\mathcal{E}=d P$ while for the branch $sT+m\vec u^2=0$ it imposes the relation $m = - \frac{d+1}{\vec u^2}P$. In addition, the number of first order transport coefficients reduces from 12 to 8 (see Appendix~\hyperref[app:dissipation]{B}). It would be interesting to consider this case in further detail and explore its connections to flat space holography.\\
	
	\textbf{Acknowledgements}
	We are grateful to Arjun Bagchi, Jan de Boer, José Figueroa-O'Farrill, Jelle Hartong, Akash Jain, Gerben Oling, Niels Obers, Stefan Prohazka, Ashish Shukla and Amos Yarom for useful discussions. The work of JA is partly
	supported by the Dutch Institute for Emergent Phenomena (DIEP) cluster at the University of Amsterdam via the programme Foundations and Applications of Emergence (FAEME). The work of EH was supported by Jelle Hartong's Royal Society University Research Fellowship (renewal) ``Non-Lorentzian String Theory'' (URF\textbackslash R\textbackslash 221038) via an enhancement award.

\providecommand{\href}[2]{#2}\begingroup\raggedright\endgroup

	\appendix
	
	\section{Appendix A: Ho\v rava--Lifshitz action for the boost Goldstone}
	\label{app:HL}
	If the Carrollian boost symmetry is spontaneously broken the associated Goldstone field is a spatial vector $\mathcal{\theta}^\mu$. It is possible to write down a low energy effective action for the boost Goldstone as in \cite{Nicolis:2015sra} coupled to a {weak} Carrollian geometry. The most general action for the Goldstone up to two derivative order is given by
	\begin{equation} \label{eq:HL}
		\begin{split}
			S_\theta=\int d^{d+1}x e~\Big[&\Lambda+c_1 \mathcal{R}+c_2\hat h^{\mu\nu}\pounds_v \hat\tau_\mu\pounds_v \hat\tau_\nu\\
			&c_3\hat h^{\mu\rho}\hat h^{\nu\sigma}K_{\mu\nu}K_{\rho\sigma}+c_4 (\hat h^{\mu\nu}K_{\mu\nu})^2\big]\,,
		\end{split}
	\end{equation}
	where $\Lambda, c_1, c_2, c_3, c_4$ are arbitrary constants and $\mathcal{R}$ is the Ricci scalar $\mathcal{R}=\hat h^{\mu\nu} \hat h^{\lambda \rho} R_{\mu\lambda\nu\rho}$ where the Riemann tensor $R_{\mu\lambda\nu\rho}$ is given in terms of the effective Aristotelian connection
	\begin{equation}
		\begin{split}
			\hat\Gamma^\rho_{\mu\nu}=&-v^\rho\partial_\mu\hat\tau_\nu+\frac{1}{2}\hat h^{\rho\sigma}\left(2\partial_{(\mu} h_{\nu)\sigma}-\partial_\sigma h_{\mu\nu}\right) \\
			&-\hat h^{\rho\sigma}\hat\tau_\nu K_{\mu\sigma}\,,
		\end{split}
	\end{equation}
	in the usual way. The action \eqref{eq:HL} is a Ho\v rava--Lifshitz-type action \cite{Hartong:2015xda, Hartong:2015zia} but only the Goldstone $\theta^\mu$ is dynamical. The energy-momentum tensor that arises from~\eqref{eq:HL} contains a constant pressure term proportional to the ``cosmological constant'' $\Lambda$ and many second order derivative terms acting on $\hat \tau_\mu$ or on $\theta_\mu$. Focusing on the case in which $\hat \tau_\mu$ acquires the expectation value $\langle\hat\tau_\mu\rangle=\delta_\mu^t$ with $\vec\theta_\mu=0$, the energy-momentum tensor is just ${\tilde T^\mu}{}_\nu=\Lambda \delta^\mu_\nu$ and hence defining the energy density $\mathcal{E}=v^\mu\hat\tau_\mu {\tilde T^\mu}{}_\nu$ we obtain the framid condition of \cite{Nicolis:2015sra, Alberte:2020eil}, namely that $\mathcal{E}+P=0$ where $P=\Lambda$.
	
	\section{Appendix B: Dissipative dynamics of Carrollian fluids}
	\label{app:dissipation}
	In this appendix we provide the details of how to go beyond ideal order dynamics and include dissipative effects in Carrollian fluids. In order to do so we take advantage of the underlying effective Aristotelian geometry, which we now describe in more detail. \\
	
	\textbf{Aristotelian action and currents.} \label{sec:Aristotelian}
	Using the Aristotelian structure, we can also write the variation~\eqref{eq:fluid-functional-variation} as
	\begin{equation}\label{eq:varAristos}
		\delta S = \int d^{d+1}x\,e\left[ - T^\mu \delta \hat\tau_\mu + \frac{1}{2}\tilde{\mathcal{T}}^{\mu\nu}\delta h_{\mu\nu} - \mathcal{K}_\mu \delta \theta^\mu\right]\,,
	\end{equation}
	where quantities carrying a ``tilde'' and $\mathcal{K}^\mu$ are Aristotelian stresses related to their Carrollian counterparts via
	\begin{equation}
		\label{eq:EMT-relation}
		\tilde{\mathcal{T}}^{\mu\nu} = \mathcal{T}^{\mu\nu} + T^\mu \theta^\nu + T^\nu\theta^\mu\,,\qquad \mathcal{K}_\mu = K_\mu - T^\nu h_{\nu\mu}\,.
	\end{equation}
	The Stueckelberg and boost Ward identities together imply that $\mathcal{K}_\mu = 0$ off-shell. We note that, due to~\eqref{eq:EMT-boost-trafo}, the modified stress-momentum tensor $\tilde{\mathcal{T}}^{\mu\nu}$ is boost invariant. It is not, however,  invariant under Stueckelberg transformations, since $\delta_S \tilde{\mathcal{T}}^{\mu\nu} = -2\chi v^\mu v^\nu \tau_\rho T^\rho$. This is expected since, due to the identity $v^\mu v^\nu \delta h_{\mu\nu} = 0$, it is only defined up to terms of the form $X v^\mu v^\mu $, where $X$ is an arbitrary function. The EMT is also invariant under both boosts and Stueckelberg transformations. When $K_\mu = 0$, the boost Ward identity $T^\nu h_{\nu\mu} = 0$ implies that $\tilde T^\mu{_\nu} =  T^\mu{_\nu} =:  -\tau_\nu T^\mu + \mathcal{T}^{\mu\rho}h_{\rho\nu}$. In the Aristotelian picture, the diffeomorphism Ward identity is given by
	\begin{equation}
		\label{eq:diffeo-WI-Aristotle}
		\begin{split}
			e^{-1}\D_\mu(e  \tilde T^\mu{_\rho}) +  T^\mu \D_\rho \hat \tau_\mu - \frac{1}{2}{\tilde{\mathcal{T}}}^{\mu\nu}\D_\rho h_{\mu\nu}=0\,.
		\end{split}
	\end{equation}
	Using the variational formulae,
	\begin{equation}
		\begin{split}
			\delta h^{\mu\nu} &= 2v^{(\mu} h^{\nu)\rho}\delta\tau_\rho - h^{\mu\rho}h^{\nu\sigma}\delta h_{\rho\sigma}\,,\\
			\delta v^\mu &= v^\mu v^\nu \delta \tau_\nu - h^{\mu\nu} v^\rho \delta h_{\rho\nu}\,,
		\end{split}
	\end{equation}
	and the variation \eqref{eq:varAristos} together with \eqref{eq:invariants}, the ideal order boost-invariant energy-momentum tensor (EMT) is
	\begin{equation} \label{eq:Aristosemt}
		\tilde T^\mu_{(0)\nu} = P\delta^\mu_\nu - (sT + m\vec u^2) u^\mu \hat \tau_\nu + m u^\mu \vec u_\nu\,.
	\end{equation}
	The relation between the EMTs is given in~\eqref{eq:EMT-relation} and reproduces~\eqref{eq:ideal-EMT}.\\
	
	\textbf{Entropy production.}
	Given the Aristotelian structure, it is straighforward to study the near-equilibrium dynamucs of Carrollian fluids. In particular, we can use the adiabaticity equation for Aristotelian fluids~\cite{Armas:2020mpr} but with the additional Goldstone field
	\begin{equation*}
		e^{-1}\D_\mu (eN^\mu) = - T^\mu \delta_{\mathscr B}\hat\tau_\mu + \frac{1}{2}\tilde{\mathcal{T}}^{\mu\nu}\delta_{\mathscr B} h_{\mu\nu} - \mathcal{K}_\mu \delta_{\mathscr B}\theta^\mu + \Delta\,,
	\end{equation*}
	where $\Delta \geq 0$, and where $\mathscr B = (\beta^\mu, \lambda^B_\mu, \chi^B)$ are a out-of-equilibrium parameters that in equilibrium reduce to the symmetry parameters $K =(k^\mu,\lambda^K_\mu,\chi^K)$. The parameters $\mathscr B$ act on the various fields according to the first equality in \eqref{eq:equil-transformations} with $K$ replaced by $\mathscr B$. Defining the free energy current $N^\mu = S^\mu + \tilde T^\mu{_\rho}\frac{u^\rho}{T}$, we can write the adiabaticity equation as the second law of thermodynamics
	\begin{equation} \label{eq:entropy}
		\begin{split}
			& e^{-1}\D_\mu (e S^\mu) \\
			&+ \frac{u^\rho}{T}\left(  e^{-1}\D_\mu(e \tilde T^\mu{_\rho}) + T^\mu \D_\rho \hat\tau_\mu - \frac{1}{2}\tilde{\mathcal{T}}^{\mu\nu}\D_\rho h_{\mu\nu} \right)\\
			&+ ({K}_\mu - T^\nu h_{\nu\mu}) \delta_{\mathscr B}\theta^\mu = \Delta\geq 0\,.
		\end{split}
	\end{equation}
	Using the equilibrium currents \eqref{eq:Aristosemt} we find that for \eqref{eq:entropy} to be satisfied we must have that
	\begin{equation}
		\label{eq:Goldstone-out-of-eq}
		K_\mu = a_{\mu\nu}\delta_{\mathscr B} \theta^\nu + T^\nu h_{\nu\mu} + \mathcal{O}(\D)\,,
	\end{equation}
	for some positive definite $a_{\mu\nu}$. When $K_\mu = 0$ and the boost Ward identity is satisfied, $T^\nu h_{\nu\mu}=0$, we therefore find that $\delta_{\mathscr B} \theta^\nu = 0$ out of equilibrium at ideal order. Given the definition of $\delta_{\mathscr B} \theta^\nu$ (see Eq.~\eqref{eq:equil-transformations} with the subscript $K$ replaced by $\mathscr B$), including corrections of $\mathcal{O}(\D)$ will allow us to determine $\lambda^B_\mu, \chi^B$ in terms of the other hydrodynamic variables and hence be eliminated from the theory. In addition, $\mathcal{O}(\D^2)$ corrections to \eqref{eq:Goldstone-out-of-eq} can be absorbed by field redefinitions of the parameters, in particular $\lambda_\mu^B \rightarrow \lambda_\mu^B + \delta \lambda_\mu^B$ to set the spatial corrections to zero and $\chi^B \rightarrow \chi^B + \delta \chi^B$ to set the temporal corrections to zero. Once such choices are implemented, Eq.~\eqref{eq:Goldstone-out-of-eq} becomes $K_\mu=T^\nu h_{\nu\mu}$ out of equilibrium, as advertised in the main text. \\
	
	\textbf{First order corrections.} At order $\mathcal{O}(\D)$ we can use the second law \eqref{eq:entropy} to constrain possible corrections to the stress-tensor of the Carrollian fluid. These corrections have been obtained in \cite{deBoer:2020xlc, Armas:2020mpr}. In particular, there are two hydrostatic corrections that can be included in the equilibrium partition function according to
	\begin{equation}
		\begin{split}
			S = \int d^{d+1}x\,e\big[P(T,\vec u^2)&-F_1(T,\vec u^2)v^\mu\partial_\mu T \\
			&\quad-F_2(T,\vec u^2)v^\mu\partial_\mu \vec u^2\big]\,,
		\end{split}
	\end{equation}
	for some arbitrary functions $F_1$ and $F_2$.
	The corrections to the EMT arising from here are explicitly computed in \cite{deBoer:2020xlc,Armas:2020mpr}. On the other hand the non-dissipative non-hydrostatic sector is trivial once Onsager's relations are imposed \cite{deBoer:2020xlc,Armas:2020mpr}. In turn, the dissipative sector in the Landau frame, where $\tilde T^\mu_{(1)} = \tilde{\mathcal{T}}^{\mu\nu}_{(1)} \vec u_\nu$, can be parametrized as
	\begin{equation}
		\label{eq:first-order-EMT}
		\tilde T^\mu_{(1)\nu} = \frac{1}{2}\left[-\eta^{\mu\rho\kappa\lambda}\vec u_\rho\hat\tau_\nu +\eta^{\mu\rho\kappa\lambda}h_{\rho\nu} \right]X_{\kappa\lambda}\,,
	\end{equation}
	where $X_{\kappa\lambda} := \pounds_u h_{\kappa\lambda} - \vec u_\kappa \pounds_u\hat\tau_\lambda - \vec u_\lambda \pounds_u \hat\tau_\kappa$, and where the $\eta$-tensor, in addition to being symmetric in each pair of indices, is symmetric under interchange of pairs of indices, i.e., $\eta^{\mu\nu\rho\sigma} = \eta^{\nu\mu\rho\sigma} = \eta^{\mu\nu\sigma\rho} = \eta^{\rho\sigma\mu\nu}$. Explicitly, the $\eta$-tensor is given by
	\begin{widetext}
		\begin{equation}\label{eq:dissipativestress}
			\begin{split}
				\eta^{\mu\nu\rho\sigma} & =  \mathfrak{t}\left(\hat P^{\mu\rho}\hat P^{\nu\sigma}+\hat P^{\mu\sigma}\hat P^{\nu\rho}-\frac{2}{d-1}\hat P^{\mu\nu}\hat P^{\rho\sigma}\right)+ \frac{4s_1}{u^2}v^{(\mu}n^{\nu)}n^{(\rho}v^{\sigma)}+s_2n^\mu n^\nu n^\rho n^\sigma+s_3\hat P^{\mu\nu}\hat P^{\rho\sigma}\\
				&\quad+\frac{4f_1}{u^2}v^{(\mu} \hat P^{\nu)(\rho} v^{\sigma)}+f_2\left(\hat P^{\mu\rho}n^\nu n^\sigma+\hat P^{\nu\rho}n^\mu n^\sigma+\hat P^{\mu\sigma}n^\nu n^\rho+\hat P^{\nu\sigma}n^\mu n^\rho\right)\\
				&\quad+s_6\left(\hat P^{\rho\sigma}n^\mu n^\nu+ \hat P^{\mu\nu}n^\rho n^\sigma\right)
				-\frac{4f_3}{\sqrt{u^2}}\left(v^{(\mu}\hat P^{\nu)(\rho}n^{\sigma)}+ n^{(\mu}\hat P^{\nu)(\rho}v^{\sigma)}\right)\\
				&\quad - \frac{2s_5}{\sqrt{u^2}}\left( v^{(\mu}n^{\nu)} \hat P^{\rho\sigma}+ \hat P^{\mu\nu} n^{(\rho}v^{\sigma)}\right) - \frac{2s_4}{\sqrt{u^2}}\left(v^{(\mu}n^{\nu)}n^\rho n^\sigma + n^\mu n^\nu n^{(\rho}v^{\sigma)}\right)\,, 
			\end{split}
		\end{equation}
	\end{widetext}
	where we have defined $\hat P^{\mu\nu} = \hat h^{\mu\nu} -  n^\mu  n^\nu$ and  $n^\mu = \frac{\vec u^\mu}{\verts{\vec u}}=\frac{\hat h^\mu_\nu u^\nu}{\verts{\vec u}}$ such that $h_{\mu\nu} n^\mu n^\nu = 1$ and $\hat P^{\mu\nu} n_\nu = 0$. The 10 dissipative coefficients appearing in \eqref{eq:dissipativestress} are all functions of $T$ and $\vec u^2$ and constrained according to
	\begin{equation}
		\begin{split}
			&\mathfrak{t}\le0~,~f_1,f_2\le0~,~f_1f_2-f_3^2\ge0\,,\\
			&s_1,s_2,s_3\le0~,~s_1s_2-s_4^2\ge0~,~s_1s_3-s_5^2\ge0\,,\\
			&s_3s_3-s_6^2\ge0~,\\
			&s_1s_2s_3-s_3s_4^2-s_2s_5^2+2s_4s_5s_6-s_1s_6^2\le0\,.
		\end{split}
	\end{equation}
	These coefficients can be viewed as different types of viscosities in different directions. In particular $s_3$ is the bulk viscosity and $\mathfrak{t}$ the shear viscosity.
	Thus, the class of Carrollian fluids we consider here is characterised by a total of 12 transport coefficients at first order in derivatives. We note that even though when formulated in terms of ${\tilde T^\mu}{}_\nu$ the stresses are the same as for an Aristotelian fluid, the actual Carrollian stresses are obtained via \eqref{eq:EMT-relation} and must satisfy the Ward identity $T^\mu h_\mu^\nu =0$. \\
	
	\textbf{Flat Carrollian space.} For the calculation of the linearised excitations of Carrollian fluids as described in the main text we focus on flat Carrollian space and choose Cartesian coordinates such that
	\begin{equation*}
		\tau_\mu = \delta_\mu^t\,,\quad v^\mu = -\delta^\mu_t\,,\quad h_{\mu\nu} = \delta_\mu^i \delta_\nu^i\,,\quad h^{\mu\nu} = \delta^\mu_i \delta^\nu_i\,.
	\end{equation*}
	The residual gauge transformations are those that preserve this choice, i.e., those that satisfy~\eqref{eq:equil-transformations}. Setting $\delta v^\mu = 0$ tells us that $\D_t \xi^\mu = 0$, which means that $\delta\tau_\mu = 0$ implies that $\lambda_t = 0$ and $\D_i \xi^t(x^j) + \lambda_i(x^j) = 0$. Finally, setting $\delta h_{\mu\nu}=0$ implies that $2\D_{(i}\xi_{j)} = 0$, which is solved by $\xi^i = a^i + \Lambda^i{_j}x^j$, 
	where $a^i$ is a constant translation and $\Lambda^{ij} = -\Lambda^{ji}$ a constant rotation. In summary, the infinite-dimensional isometries of flat Carrollian space are
	\begin{equation}
		\xi^t = c + f(x^i)\,,\qquad \xi^i = a^i + \Lambda^i{_j}x^j\,,
	\end{equation}
	where $c$ is a constant, and $f(x^i)$ an arbitrary function of $x^i$ without a constant part. As explained in~\cite{Duval:2014uoa}, this reduces to the generators of the Carroll algebra when we require the adapted affine connection to be preserved as well. The Goldstone transforms as $\delta\theta^\mu = \pounds_\xi \theta^\mu - h^{\mu\nu}\lambda_\nu + v^\mu \chi$, and we may fix the Stueckelberg symmetry by setting $\theta^t = 0$, which, in order to be stable under variations, requires     $\delta \theta^t = -\chi -\theta^i \D_i \xi^t = 0$, i.e., $\chi = \theta^i\lambda_i$. With these choices, the spatial components of the Goldstone transform as $\delta \theta^i = \xi^j\D_j \theta^i  - \theta^j \D_j\xi^i - \lambda_i$. Relative to the above flat Carrollian structure, the components of the fluid velocity are then $u^\mu = (1 - \theta_i v^i,v^i)$, where $h^\mu_\nu u^\nu \overset{\text{flat}}{\longrightarrow} v^i$. The components of the EMT are thus
	\begin{align}
		\begin{aligned}
			T^0{_0} &= P\,,&T^0{_i}&= m(1-{\theta_j v^j})v_i\,,\\
			T^{i}{_0} &= 0\,,& T^{i}{_j} &= P \delta^i_j + mv^i v_j\,.
		\end{aligned}
	\end{align}
	When $\theta = 0$, these are the components of the Carroll EMT in Eqs. (2.58)--(2.61) as derived in~\cite{deBoer:2017ing}.
	
	In order to obtain the components of the dissipative contributions to the EMT we note that in flat Carrollian space we have that the components of $n^\mu$ are given by $n^t = -\frac{1}{\verts{v}} \theta_i v^i$ and $n^i = \frac{1}{\verts{v}}v^i$, while the projector $\hat P^{\mu\nu}$ has components
	\begin{equation*}
		\begin{split}
			\hat P^{ij} &= P^{ij} = \delta^{ij} - n^i n^j\,,\qquad
			\hat P^{0i} = -\theta^i + \frac{v^i}{\verts{v}^2}v^j \theta_j \,,\\
			\hat P^{00} &= \theta_i \theta^i - \frac{(\theta_i v^i)^2}{\verts{v}^2}\,.
		\end{split}
	\end{equation*}
	Note in particular that $P^{ij}v^j = \hat P^{0i}v^i = 0$. In addition, $X_{\mu\nu}$ has components
	\begin{equation*}
		\begin{split}
			X_{00} &= 0\,,\qquad X_{0i} = \D_t v^i + v^i v^j \D_t\theta_j\,,\\
			X_{ij} &= 2\D_{(i}v_{j)} - 2v_{(i}\D_t\theta_{j)} (1 - \theta_k v^k) - 2v^k(v_{(i\vert}\D_k \theta_{\vert j)}\\
			&\quad+ v_{(i}\D_{j)}\theta_k) \,.
		\end{split}
	\end{equation*}
	This means that the components of the dissipative corrections to the EMT in flat Carrollian space are
	\begin{eqnarray}
		\begin{aligned}
			\tilde T^0_{(1)0} &= -\frac{1}{2}\kappa^{jki} v^i X_{jk} - \kappa^{ij}v^i X_{0j} \,,\\
			%
				\tilde T^0_{(1)i} &= \frac{1}{2}X_{kl}  ( \kappa^{kli} - \kappa^{klj}v^j\theta_i ) + X_{0l}(\kappa^{il} - \kappa^{jl} v^j\theta_i ) \,,\\
			%
			\tilde T^i_{(1)0} &= -\frac{1}{2} \eta^{ijkl} v^j X_{kl} - \kappa^{ijl}v^j X_{l0}\,,\\
			%
				\tilde T^i_{(1)j} &= \frac{1}{2}(\eta^{ijkl} - \eta^{imkl}v^m\theta_j )X_{kl} \\
				&\quad+ ( \kappa^{ijk} - \kappa^{imk}v^m \theta_j )X_{0k} \,,
			\end{aligned}
		\end{eqnarray}  
		where we defined $\kappa^{ij} = \eta^{0i0j}$ and $\kappa^{ijk} = \eta^{ijk0}$. We use these components in the main text in order to obtain the linear spectrum of excitations.\\
		
		\textbf{BMS fluids.} The $(d+1)$-dimensional BMS group and the $d$-dimensional conformal Carroll group are isomorphic, $\operatorname{CCar}_{d}\cong \operatorname{BMS}_{d+1}$~\cite{Duval:2014uva,Duval:2014lpa}. The additional conformal transformations act on the Carrollian structure and its inverse according to
		\begin{equation*}
			\begin{split}
				&\delta h_{\mu\nu} = \omega h_{\mu\nu}\,,~ \delta v^\mu = - \frac{\omega}{2}v^\mu\,,~\delta \tau_\mu = \frac{\omega}{2}\tau_\mu\,,\\
				&\delta h^{\mu\nu} = -\omega h^{\mu\nu}\,,
			\end{split}
		\end{equation*}
		where $\omega$ is the parameter of the conformal rescaling. If we demand that the effective Aristotelian structure transforms as above, we must demand that $\delta \theta^\mu = -\frac{\omega}{2}\theta^\mu$. This leads to the following additional Ward identity
		\begin{equation}
			-T^\mu \tau_\mu + \mathcal{T}^{\mu\nu}h_{\mu\nu}  = -K_\mu \theta^\mu\,,
		\end{equation}
		which is both boost and Stueckelberg invariant. This implies that the ideal order EMT~\eqref{eq:ideal-EMT} for the branch with $\vec u_\mu = 0$ becomes
		\begin{equation}
			T^\mu_{(0)\nu} = P\delta^\mu_\nu + (d+1)P v^\mu\hat\tau_\nu\,.
		\end{equation}
		In particular, we get the relation $sT = \mathcal{E} + P = (d+1)P$, implying that $\mathcal{E} = dP$. The ideal EMT for the branch with $sT + m\vec u^2 = 0$ is
		\begin{equation}
			T^\mu_{(0)\nu} = P\delta^\mu_\nu - \frac{d+1}{2\vec u^2}Pu^\mu \vec u_\nu\,.
		\end{equation}
		For this fluid, we have that $sT + m\vec u^2 = \mathcal{E} + P = 0$, while the conformal Ward identity implies the relation $m = - \frac{d+1}{\vec u^2}P$. In the language of~\cite{deBoer:2020xlc,Armas:2020mpr}, this corresponds to a $z=1$ Lifshitz fluid when expressed in terms of the effective Aristotelian structure, which was shown to reduce the number of transport coefficients at first order from 12 to 8.

		\section{Appendix C: The small-\texorpdfstring{$c$}{c} limit of relativistic fluids}
		\label{app:limit}
		In this appendix we give additional details of the $c\rightarrow 0$ limit of relativistic fluids. We begin by summarising the $c^2$-expansion of Lorentzian geometry as presented in~\cite{Hansen:2021fxi}. We first express the metric and its inverse in terms of PUL variables
		\begin{equation}
			g_{\mu\nu} = -c^2 T_\mu T_\nu + \Pi_{\mu\nu}\,,\quad g^{\mu\nu} = - \frac{1}{c^2}V^\mu V^\nu + \Pi^{\mu\nu}\,,
		\end{equation}
		which satisfy the relations
		\begin{equation}
			\begin{split}
				T_\mu V^\mu &= -1\,,\qquad T_\mu \Pi^{\mu\nu} = V^\mu \Pi_{\mu\nu} = 0\\
				\Pi^{\mu\rho}\Pi_{\rho\nu} &= \delta^\mu_\nu + V^\mu T_\nu \,.
			\end{split}
		\end{equation}
		The leading order pieces of the PUL variables correspond to the fields that make up the Carrollian structure and its inverse, i.e.,
		\begin{align}
			\label{eq:PUL-expansions}
			\begin{aligned}
				V^\mu &= v^\mu + \mathcal{O}(c^2)\,,& \Pi^{\mu\nu} &= h^{\mu\nu} + \mathcal{O}(c^2)\,,\\ T_\mu &= \tau_\mu +  \mathcal{O}(c^2)\,, &\Pi_{\mu\nu} &= h_{\mu\nu} +  \mathcal{O}(c^2)\,,
			\end{aligned}
		\end{align}
		but it is useful to work with the PUL variables and only perform the expansions in~\eqref{eq:PUL-expansions} at the end. The relativistic EMT is given by Eq.~\eqref{eq:relativistic-EMT}, where
		\begin{equation}
			\hat{\mathcal{E}} = \mathcal{E} + \mathcal{O}(c^2)\,,\qquad \hat P = P + \mathcal{O}(c^2)\,,
		\end{equation}
		while the fluid velocity $U^\mu$, which is timelike, satisfies $U^\mu U^\nu g_{\mu\nu} = -c^2$, and is given by 
		\begin{equation}
			U^\mu = -V^\mu - c^2\mathfrak u^\mu \,,
		\end{equation}
		while
		\begin{equation}
			U_\mu = g_{\mu\nu}U^\nu = -c^2 (T_\mu + t\vec{\mathfrak u}_\mu ) \,,
		\end{equation}
		where we defined $\vec{\mathfrak u}_\mu := \Pi_{\mu\nu}\mathfrak u^\nu $. This leads to
		the following EMT
		\begin{equation}
			\label{eq:PUL-EMT}
			\begin{split}
				T^\mu{_\nu} &= \hat P\delta^\mu_\nu\\
				&+ (\hat{\mathcal{E}} + \hat P)\Big[ V^\mu T_\nu + V^\mu \vec{\mathfrak u}_\nu + c^2 T_\nu \mathfrak u^\mu + c^2 \mathfrak u^\mu \vec{\mathfrak u}_\nu \Big]\,~.
			\end{split}
		\end{equation}
		The leading order piece of this EMT reproduces Eq.~\eqref{eq:EMT-from-limit}.
		
		Given the EMT in~\eqref{eq:PUL-EMT}, we can take the $c\to0$ limit of the relativistic fluid equations of motion, which read
		\begin{equation}
			\label{eq:relativistic-fluid-eq}
			\hat\nabla_\mu T^\mu{_\nu} = 0\,,
		\end{equation}
		where $\hat\nabla$ is the Levi-Civita connection, whose behaviour in the Carroll limit was studied in~\cite{Hansen:2021fxi} as we now review. The Levi-Civita connection can be decomposed as
		\begin{equation}
			\hat\Gamma^\rho_{\mu\nu} = -\frac{1}{c^2}V^\rho \mathcal{K}_{\mu\nu} + \tilde C^\rho_{\mu\nu} + S^\rho_{\mu\nu} + \mathcal{O}(c^2)\,,
		\end{equation}
		where we defined
		\begin{equation}
			\mathcal{K}_{\mu\nu} = -\frac{1}{2}\pounds_V \Pi_{\mu\nu}\,,
		\end{equation}
		which in particular satisfies
		\begin{equation}
			V^\mu \mathcal{K}_{\mu\nu} = 0\,,
		\end{equation}
		\vspace{-0.8cm}
		
		\noindent and where $\tilde C^\rho_{\mu\nu}$ (with associated derivative $\overset{(\tilde C)}{\nabla}$) satisfies 
		\vspace{-0.1cm}
		\begin{equation}
			\overset{(\tilde C)}{\nabla}_\mu V^\nu = \overset{(\tilde C)}{\nabla}_\mu \Pi_{\nu\lambda} = 0\,.
		\end{equation}
		Explicitly, it is given by
		\begin{equation}
			\begin{split}
				\tilde C^\rho_{\mu\nu} &= -V^\rho\D_{(\mu}T_{\nu)} - V^\rho T_{(\mu}\pounds_{V}T_{\nu)}\\
				&\qquad + \frac{1}{2}\Pi^{\rho\lambda}(\D_\mu \Pi_{\nu\lambda} + \D_\nu \Pi_{\mu\lambda} - \D_\lambda \Pi_{\mu\nu})\\
				&\qquad- \Pi^{\rho\lambda} T_\nu \mathcal{K}_{\mu\lambda}\,.
			\end{split}
		\end{equation}
		In turn, the ``difference tensor'' $S^\rho_{\mu\nu}$ is given by
		\begin{equation}
			S^\rho_{\mu\nu} = \Pi^{\rho\lambda} T_\nu \mathcal{K}_{\mu\lambda}\,.
		\end{equation}
		This implies that we can write 
		\begin{equation}
			\begin{split}
				\hat \nabla_\mu V^\nu &= \tilde\nabla_\mu V^\nu - \frac{1}{c^2}V^\nu\mathcal{K}_{\mu\rho}V^\rho + \Pi^{\nu\lambda}T_\rho \mathcal{K}_{\mu\lambda}V^\rho \\
				&= -\Pi^{\nu\lambda}\mathcal{K}_{\mu\lambda}\,,
			\end{split}
		\end{equation}
		which is spatial in the sense that
		\begin{equation}
			\label{eq:useful-PUL-relation}
			V^\mu \hat\nabla_\mu V^\nu = T_\nu \hat\nabla_\mu V^\nu = 0\,.
		\end{equation}
		The temporal projection of~\eqref{eq:relativistic-fluid-eq} is $V^\mu \hat\nabla_\nu T^\nu{_\mu}=0$, which at $\mathcal{O}(1)$ gives rise to the following expression in PUL variables 
		\begin{equation*}
			\begin{split}
				V^\mu \D_\mu \hat{\mathcal{E}} &= - (\hat{\mathcal{E}} + \hat P)\left[ \hat\nabla_\nu V^\nu - V^\nu T_\mu \hat \nabla_\nu V^\mu - \vec{\mathfrak u}_\mu V^\nu\hat\nabla_\nu V^\mu  \right]\\
				&=- (\hat{\mathcal{E}} + \hat P) \hat\nabla_\nu V^\nu\,,
			\end{split}
		\end{equation*}
		where we used~\eqref{eq:useful-PUL-relation}. In the $c \rightarrow 0$ limit, this gives rise to the equation
		\begin{equation}
			\label{eq:temporal-proj}
			v^\mu \D_\mu \mathcal{E} = ({\mathcal{E}} + P) K\,,
		\end{equation}
		where
		\begin{equation}
			K = h^{\mu\nu} K_{\mu\nu} = \Pi^{\mu\nu}\mathcal{K}_{\mu\nu}\big\vert_{c=0}\,,
		\end{equation}
		is the trace of the intrinsic torsion of thte Carrollian structure. The spatial projection of the equation of motion is $\Pi^{\lambda\mu}\hat\nabla_\nu T^\nu{_\mu} = 0$, which gives
		\begin{equation*}
			\begin{split}
				\Pi^{\lambda\mu}\D_\mu \hat P &= -(\hat{\mathcal{E}} + \hat P)\Pi^{\lambda\mu}V^\nu\hat\nabla_\nu T_\mu + (\hat{\mathcal{E}} + \hat P)\mathcal{K} \Pi^{\lambda\sigma}\vec{\mathfrak u}_\sigma\\
				&\hspace{-0.5cm} - \Pi^{\lambda\mu} V^\nu \hat\nabla_\nu((\hat{\mathcal{E}} + \hat P)\vec{\mathfrak u}_\mu) + (\hat{\mathcal{E}} + \hat P) \Pi^{\lambda\mu} \Pi^{\nu\sigma} \vec{\mathfrak u}_\sigma \mathcal{K}_{\mu\nu} \,.
			\end{split}
		\end{equation*}
		Hence, if we define
		\begin{equation}
			\begin{split}
				\widetilde\varphi^\lambda&:=\left( \Pi^{\lambda\mu}V^\nu\hat\nabla_\nu T_\mu - \Pi^{\lambda\mu} \Pi^{\nu\sigma} \vec{\mathfrak u}_\sigma \mathcal{K}_{\mu\nu}\right)\bigg\vert_{c=0}\\
				&~=h^{\lambda \mu}v^\nu \tau_{\nu\mu} - h^{\lambda\mu}h^{\nu\sigma}\vec{\theta}_\sigma K_{\mu\nu}\,,
			\end{split}
		\end{equation}
		the spatial part of the equation in the $c\rightarrow 0$ limit reduces to
		\begin{equation}
			\label{eq:spatial-proj}
			\begin{split}
				h^{\lambda\mu}\D_\mu P &= -({\mathcal{E}} + P)\widetilde\varphi^\lambda + ({\mathcal{E}} + P){K} h^{\lambda\sigma} \vec{\theta}_\sigma\\
				&\quad - h^{\lambda\mu} v^\nu \nabla_\nu^{(0)}(({\mathcal{E}} + P)\vec{\theta}_\mu)\\
				&= -({\mathcal{E}} + P)\widetilde\varphi^\lambda + ({\mathcal{E}} + P){K} h^{\lambda\sigma} \vec{\theta}_\sigma\\
				&\quad - h^{\lambda\mu} v^\nu \tilde\nabla_\nu(({\mathcal{E}} + P)\vec{\theta}_\mu)
				\,,
			\end{split}
		\end{equation}
		where $\nabla_\mu^{(0)}$ involves the connection 
		\begin{equation}
			\Gamma^{(0)\rho}_{\mu\nu} = \big[\tilde C^\rho_{\mu\nu} + S^\rho_{\mu\nu}\big]_{c=0} = \tilde\Gamma^\rho_{\mu\nu} + h^{\rho\lambda}\tau_\nu K_{\mu\lambda}\,,
		\end{equation}
		and which is torsion-free, i.e., ${\Gamma}^{(0)\rho}_{[\mu\nu]} = 0$. The connection $\tilde\Gamma^\rho_{\mu\nu} = \tilde C^\rho_{\mu\nu}\big\vert_{c=0}$ with associated covariant derivative $\tilde\nabla = \overset{(\tilde C)}{\nabla}\big\vert_{c=0}$ is compatible with the Carrollian structure, i.e., 
		\begin{equation}
			\tilde\nabla_\mu v^\nu = \tilde \nabla_\mu h_{\nu\rho} = 0\,.
		\end{equation}
		Equations~\eqref{eq:temporal-proj} and~\eqref{eq:spatial-proj} were presented as Eq.~\eqref{eq:limit-equations} in the main text.

		At the same time, we can rewrite the equation of motion~\eqref{eq:diffeo-WI-Carroll} to bring it to the form above. In particular, we may rewrite~\eqref{eq:diffeo-WI-Carroll} for \textit{any} connection $\mathscr D$ as \cite{deBoer:2020xlc}
		\begin{equation}
			\label{eq:eom-generic}
			\begin{split}
				&\mathscr D_\mu T^\mu{_\nu} + T^\mu \mathscr D_\nu \tau_\mu - \frac{1}{2}{\mathcal{T}}^{\mu\rho}\mathscr D_\nu h_{\mu\rho}\\
				&\quad - (\Upsilon^\mu_{\mu\rho} - e^{-1}\D_\rho e)T^\rho{_\nu} + 2\Upsilon^\sigma_{[\mu\nu]}T^\mu{_\sigma} = 0 \,,
			\end{split}
		\end{equation}
		where $\Upsilon^\rho_{\mu\nu}$ are the connection components of $\mathscr D$. Setting $\mathscr D = \nabla^{(0)}$, we first note the following useful identities
		\begin{equation}
			\begin{split}
				\nabla^{(0)}_\mu v^\nu &= - h^{\nu\lambda} K_{\mu\lambda}\,,\\
				e^{-1}\D_\mu e &= \tilde\Gamma^\nu_{\nu\mu} + \tau_\mu K = \Gamma^{(0)\nu}_{\nu\mu}\,,\\
				\nabla_\mu^{(0)}\tau_\nu &= \tilde\nabla_\mu \tau_\nu = \frac{1}{2}\tau_{\mu\nu} - v^\rho \tau_{\rho(\mu}\tau_{\nu)}\,,\\
				\nabla^{(0)}_\mu h_{\nu\rho} &= -2K_{\mu(\nu}\tau_{\rho)}\,,\\
				\nabla^{(0)}_\mu h^{\nu\rho} &= \tilde\nabla_\mu h^{\nu\rho} = -v^{(\nu}h^{\rho)\sigma}\tau_{\sigma\lambda}(\delta^\lambda_\mu - v^\lambda\tau_\mu)\,,
			\end{split}
		\end{equation}
		where we defined the ``field strength'' $2$-form of $\tau$ as
		\begin{equation}
			\tau_{\mu\nu} = 2\D_{[\mu}\tau_{\nu]}\,.
		\end{equation}
		The EMT of the ``timelike'' Carrollian fluid is
		\begin{equation}
			T^\mu{_\nu} = P\delta^\mu_\nu + ({\mathcal{E}} + P)v^\mu (\tau_\nu - \vec \theta_\nu)\,,
		\end{equation}
		while the energy current and momentum-stress tensor are
		\begin{equation}
			\begin{split}
				T^\mu &=- \tilde{\mathcal{E}} v^\mu\,,\\
				\mathcal{T}^{\mu\nu} &= P h^{\mu\nu} - 2(\tilde{\mathcal{E}} + P)v^{(\mu} h^{\nu)\rho}\vec\theta_{\rho}\,.
			\end{split}
		\end{equation}
		Using the connection $\nabla^{(0)}$, the equation of motion~\eqref{eq:eom-generic} becomes
		\begin{equation}
			\nabla_\mu^{(0)}T^{\mu}{_\nu} + T^\mu \left( \frac{1}{2}\tau_{\nu\mu} - v^\rho \tau_{\rho(\mu}\tau_{\nu)} \right) + \mathcal{T}^{\mu\rho} K_{\nu\mu}\tau_\rho = 0\,.
		\end{equation}
		The temporal projection of this equation gives~\eqref{eq:temporal-proj}, while the spatial projection recovers~\eqref{eq:spatial-proj}. \\
		
		\textbf{``Randers--Papapetrou form.''} We can explicitly match the constitutive relations and the equations of motion with those that appear in~\cite{Petkou:2022bmz,Bagchi:2023ysc}, where it is assumed that we are in a set of coordinates $x^\mu = (t,\vec x)$ such that the Carrollian structure is of the form
		\begin{equation}
			v^\mu \D_\mu = - \frac{1}{\Omega(t,\vec x)}\D_t\,,\qquad h_{\mu\nu}dx^\mu dx^\nu = a_{ij}(t,\vec x)dx^idx^j\,,
		\end{equation}
		while the inverse timelike vielbein $\tau$ takes the form
		\begin{equation}
			\tau_\mu dx^\mu = \Omega(t,\vec x) dt - b_i(t,\vec x) dx^i\,.
		\end{equation}
		Given these, the completeness relation $h^{\mu\rho}h_{\rho\nu} = \delta^\mu_\nu + \tau_\nu v^\mu$ and the relation $h^{\mu\nu}\tau_\nu = 0$ imply that, relative to the coordinates $(t,\vec x)$, the object $h^{\mu\nu}\D_\mu \D_\nu$ has components
		\begin{equation}
			h^{ij} = a^{ij}\,,\qquad h^{0i} = a^{ij}\frac{b_j}{\Omega}\,,\qquad h^{00} = a^{ij}\frac{b_i b_j}{\Omega^2}\,,
		\end{equation}
		where $a^{ik} a_{kj} =\delta^i_j$. In turn, this implies that $h^\mu_\nu$ has components
		\begin{equation}
			h^i_j = \delta^i_j\,,\qquad h^0_i = \frac{b_i}{\Omega}\,,\qquad h^i_0 = 0\,,
		\end{equation}
		and that the intrinsic torsion has components
		\begin{equation}
			K_{ij} = \frac{1}{2\Omega}\D_t a_{ij}\,, 
		\end{equation}
		implying that its trace is 
		\begin{equation}
			K = h^{\mu\nu}K_{\mu\nu} = \frac{1}{2\Omega}a^{ij}\D_t a_{ij}\,.
		\end{equation}
		Using Jacobi's formula for the derivative of the determinant of an invertible matrix $\D_t a = a\, a^{ij}\D_t a_{ij}$, where $a = \det(a)$, we may write
		\begin{equation}
			K = \frac{1}{\Omega}\D_t \log \sqrt{a}\,,
		\end{equation}
		which is the form appearing in~\cite{Petkou:2022bmz,Bagchi:2023ysc}. The relevant components of the $2$-form $d\tau$ are
		\begin{equation}
			v^\mu\tau_{\mu \nu}h^\nu_i = \Omega^{-1}(\D_t b_i + \D_i \Omega)\,.
		\end{equation}
		Finally, we note that 
		\begin{equation}
			-v^\nu \tilde\nabla_\nu \vec w_\mu = \Omega^{-1}\D_t \vec w_\mu - \frac{1}{2\Omega} \delta^i_\mu a^{jk}\D_t a_{ik}\vec w_j\,,
		\end{equation}
		for a spatial $1$-form $\vec w_\mu$ satsfying $v^\mu\vec w_\mu = 0$. The temporal equation in these coordinates therefore takes the form
		\begin{equation}
			\D_t \mathcal{E} = -(\mathcal{E} + P)\D_t \log \sqrt{a}\,,
		\end{equation}
		in agreement with~\cite{Petkou:2022bmz,Bagchi:2023ysc}. By lowering the index of the spatial equation~\eqref{eq:spatial-proj} using $h_{\mu\lambda}$, it becomes
		\begin{equation}
			\begin{split}
				h_{\mu}^\nu\D_\nu P &= -({\mathcal{E}} + P)\widetilde\varphi^\nu h_{\nu\mu} + ({\mathcal{E}} + P){K} h^{\nu}_\mu \vec{\theta}_\nu\\
				&\quad - h^{\lambda}_\mu v^\nu \tilde\nabla_\nu(({\mathcal{E}} + P)\vec{\theta}_\lambda)\,,    
			\end{split}
		\end{equation}
		which reduces to
		\begin{equation}
			\begin{split}
				\left(\D_i + \frac{b_i}{\Omega}\D_t\right) P &= -(\mathcal{E} + P)\Omega^{-1}(\D_t b_i + \D_i \Omega)\\
				&\quad + \left[\D_t + \frac{1}{\Omega}(\D_t \log \sqrt{a})\right] \big((\mathcal{E} + P) \vec\theta_i \big)\,,
			\end{split}
		\end{equation}
		in agreement with the spatial equation of motion derived in~\cite{Petkou:2022bmz,Bagchi:2023ysc} once we identity $\vec \theta_i = -\beta_i$.

		\section{Appendix D: Gauging the Carroll algebra and more general classes of Carrollian fluids}
		\label{app:strong-fluids}
		
		In this appendix, we construct (strong) Carrollian geometry as a Cartan geometry by ``gauging'' the Carroll algebra following~\cite{Hartong:2015xda,Figueroa-OFarrill:2022mcy} (see also~\cite{Figueroa-OFarrill:2022nui}). In particular, the adapted affine connection is not given solely in terms of the Carrollian structure and its inverse, but involves the boost component of the Ehresmann connection. From a hydrodynamic point of view, this boost component can be used to build gauge-invariant scalars in a way that is reminiscent of fracton superfluids~\cite{Armas:2023ouk,Jain:2023nbf}.\\ 
		
		\textbf{Gauging the Carroll algebra and the affine connection.}
		The Carroll algebra $\mathfrak c$ is generated by temporospatial translations $(H,P_a)$, rotations $L_{ab}$ and Carroll boosts $C_a$ with brackets given by
		\begin{equation}
			\begin{split}
				[P_a,C_b] &= \delta_{ab}H\,,\\
				[L_{ab},P_c] &= \delta_{ac}P_b - \delta_{bc}P_a\,,\\
				[L_{ab},C_c] &= \delta_{ac}C_b - \delta_{bc}C_a\,,\\
				[L_{ab},L_{cd}] &= \delta_{ac}L_{bd} - \delta_{bc}L_{ad} + \delta_{bd}L_{ac} - \delta_{ad} L_{bc}\,.
			\end{split}
		\end{equation}
		Flat Carrollian spacetime $C$ is a reductive homogeneous space of the Carroll group, which infinitesimally is captured by the Klein pair $(\mathfrak c,\mathfrak h)$, where the stabiliser is $\h \cong \braket{L_{ab},C_a}$. Being reductive, we may write $\mathfrak c = \m \oplus \h$, where $\m = \braket{H,P_a}$ satisfies $[\h,\m] \subset \m$. We may then construct a Cartan geometry on a manifold $M$ modelled on the Klein geometry $C$ by writing down a $\mathfrak c$-valued $1$-form on $M$ known as the ``Cartan connection''
		\begin{equation}
			\label{eq:cartan-connection}
			A_\mu = \tau_\mu H + e_\mu^a P_a + \psi_\mu^a C_a + \omega_\mu^{ab}L_{ab}\,,
		\end{equation}
		where the $\m$-valued fields $(\tau_\mu,e_\mu^a)$ define a coframe on $M$, with an inverse frame $(v^\mu,e^\mu_a)$ defined by the relations
		\begin{equation*}
			v^\mu \tau_\mu = -1\,,\quad v^\mu e_\mu^a = \tau_\mu e^\mu_a = 0\,,\quad -v^\mu\tau_\nu + e^\mu_a e^a_\nu = \delta^\mu_\nu\,.
		\end{equation*}
		Under $\h$-gauge transformations, $A$ transforms as
		\begin{equation}
			\delta A = d\Lambda + [A,\Lambda]\,,
		\end{equation}
		where $\Lambda$ is an $\h$-valued function parameterising the gauge transformation, which we write as
		\begin{equation}
			\Lambda = \frac{1}{2}O^{ab}L_{ab} + \lambda^a C_a\,,
		\end{equation}
		where $O^{ab}$ parameterises infinitesimal rotations and $\lambda^a$ parameterises infinitesimal Carroll boosts. This means that the gauge fields appearing in the Cartan connection~\eqref{eq:cartan-connection} transform as
		\begin{equation}
			\begin{split}
				\delta \tau_\mu &= e_\mu^a\lambda_a\,,\\
				\delta e^a_\mu &= O^a{_b}e_\mu^b\,,\\
				\delta\psi_\mu^a &= \D_\mu \lambda^a + O^a{_b}\psi^b_\mu - \lambda_b \omega_\mu^{ab}\,,\\
				\delta \omega_\mu^{ab} &= \D_\mu O^{ab} + O^a{_c}\omega_\mu^{cb} - O^b{_c}\omega_\mu^{ca}\,.
			\end{split}
		\end{equation}
		Together, $(v^\mu, e_\mu^a)$ define the Carrollian structure. The $\h$-valued fields $(\psi_\mu^a,\omega_\mu^{ab})$ together constitute an Ehresmann connection. This Ehresmann connection induces an adapted affine connection on the tangent bundle, but, unlike for (pseudo-)Riemannian gemoetry, the affine connection $\Gamma^\rho_{\mu\nu}$ cannot in general be expressed solely in terms of the Carrollian structure: it involves the boost connection, as demonstrated in~\cite{Hartong:2015xda} (see also~\cite{Bekaert:2015xua}). Since the Carrollian structure is supplemented with an adapted connection, the gauging procedure produces a strong Carrollian geometry. It is useful to define the combination
		\begin{equation}
			\label{eq:trafo-of-boost-connection}
			\varpi_{\mu\nu} := \psi_{\mu a} e^a_\nu\,,
		\end{equation}
		which transforms as
		\begin{equation}
			\delta_C \varpi_{\mu\nu} = \nabla_\mu \lambda_\nu\,,
		\end{equation}
		under Carroll boosts. The components of the affine connection $\nabla$ are explicitly given by~\cite{Hartong:2015xda}
		\begin{equation}
			\label{eq:carroll-affine-connection}
			\begin{split}
				\Gamma^\rho_{\mu\nu} &= -v^\rho \D_\mu \tau_\nu + \frac{1}{2}h^{\rho\lambda}(\D_\mu h_{\lambda\nu} + \D_\nu h_{\lambda\mu} - \D_\lambda h_{\mu\nu})\\
				&\quad - h^{\rho\lambda}\tau_\nu K_{\mu\lambda} + v^\rho \varpi_{\mu\nu} - Y^\rho{_{\mu\nu}} \,,
			\end{split}
		\end{equation}
		where $Y^\rho{_{\mu\nu}}$ is a tensor that explicitly depends on the rotation connection $\omega_\mu^{ab}$. Note that this in particular implies that the right hand side of~\eqref{eq:trafo-of-boost-connection} is linear in $\varpi_{\mu\nu}$. Finally, we remark that the adapted connection~\eqref{eq:carroll-affine-connection} is only defined up to terms of the form $v^\rho \Sigma_{\mu\nu}$, where $v^\mu \Sigma_{\mu\nu} = 0$ and $\Sigma_{\mu\nu} = \Sigma_{(\mu\nu)}$. The simplest choice is to set $\Sigma_{\mu\nu} = 0$. \\
		
		\textbf{``Strong'' Carrollian fluids \textit{\&} fractons.} So far we have described what one might call ``weak'' Carrollian fluids, since they are defined on a weak Carrollian geometry without an adapted connection. As we saw above, the adapted connection~\eqref{eq:carroll-affine-connection} involves the Ehresmann connection $(\varpi_{\mu\nu},\omega_\mu^{ab})$, where $\varpi_{\mu\nu}$ transforms under boosts according to~\eqref{eq:trafo-of-boost-connection}, while the rotation connection is boost invariant. Using the boost Goldstone $\theta^\mu$, we may form new gauge-invariant scalars. Although an analysis of such scalars is beyond the scope of this work, let us point out that the variation of the corresponding fluid functional takes the form
		\begin{equation*}
			\hspace{-0.8cm}
			\begin{split}
				\delta S &= \int d^{d+1}x\,e\Big[ - T^\mu \delta \tau_\mu + \frac{1}{2}\mathcal{T}^{\mu\nu}\delta h_{\mu\nu} + \frac{1}{2}\mathcal{S}^\mu{_{ab}}\delta\omega_\mu^{ab} \\
				&\quad \qquad\qquad\qquad\qquad\qquad + J^{\mu\nu}\delta \varpi_{\mu\nu} - K_\mu \delta \theta^\mu\Big]\,,
			\end{split}
		\end{equation*}
		where $\mathcal{S}^\mu{_{ab}}$ is the spin current (see, e.g.,~\cite{Armas:2014rva,Jain:2018jxj,Gallegos:2021bzp}), while $J^{\mu\nu}$ a the Carroll boost current. The boost Ward identity is now $T^\nu h_{\nu\mu} + h_{\mu\rho}\nabla_\nu J^{\nu\rho} = K_\mu$, which bears a striking resemblance to the dipole Ward identity for fracton superfluids~\cite{Armas:2023ouk,Jain:2023nbf}. Given the particle Carroll/fracton duality~\cite{Figueroa-OFarrill:2023vbj,Figueroa-OFarrill:2023qty}, we plan to revisit this relation between the two theories in the future.
		
	\end{document}